\documentclass{epl}
\usepackage{amsmath}

\def\Xint#1{\mathchoice
{\XXint\displaystyle\textstyle{#1}}%
{\XXint\textstyle\scriptstyle{#1}}%
{\XXint\scriptstyle\scriptscriptstyle{#1}}%
{\XXint\scriptscriptstyle\scriptscriptstyle{#1}}%
\!\int}
\def\XXint#1#2#3{{\setbox0=\hbox{$#1{#2#3}{\int}$}
\vcenter{\hbox{$#2#3$}}\kern-.63\wd0}}

\def\dashint{\Xint-}

\newcommand{\mur}{\mu_{\mathrm r}}
\newcommand{\eps}{\epsilon}

\title{Local amplitude equation from non-local dynamics}
\shorttitle{Local amplitude equation}

\author{Ren\'{e} Friedrichs $^{1,2}$ \and Andreas Engel $^1$}
\institute{$^1$ITP, Otto-von-Guericke-Universit\"{a}t,
             Postfach 4120,
             D-39016 Magdeburg, Germany\\[1ex]
           $^2$ABB AG, Corporate Research Center,
            Wallstadter Str. 59, D-68526 Ladenburg, Germany}

\begin{document}

\maketitle

\begin{abstract}
We derive a closed equation for the shape of the free surface of a
magnetic fluid subject to an external magnetic field. The equation is
strongly non-local due to the long range character of the magnetic
interaction. We develop a systematic multiple scale perturbation
expansion in which the non-locality is reduced to the occurrence of
the Hilbert transform of the surface profile. The resulting third
order amplitude equation describing the slow modulation of the basic
pattern is shown to be purely local.
\end{abstract}


\section{Introduction}

The emergence of spatio-temporal order in distributed systems can
often be theoretically analyzed in terms of amplitude equations
\cite{cross:93,ArKr}. Generically these amplitude equations are 
non-linear partial differential equations for the slow time and space
variations of an envelope function of unstable modes. As such
they reflect the {\em local} character of the underlying dynamics. 

In several interesting cases, however, the dynamics of the system is
substantially influenced by {\em non-local} interactions. These may 
arise, e.g., from a mean flow in convection problems 
\cite{SiZi,Hall,PiKn,PlBu}, from 
electric and magnetic fields in solids \cite{Elmer,Schoell} or
electrochemical systems \cite{LOMPKE}, from long range elastic
interactions \cite{kassner:02}, or from global
couplings in systems with chemical reactions
\cite{LeZo,FEN,NOWB}. 

Since the type and stability of the emerging patterns is usually
modified if non-local interactions are present the corresponding
amplitude equations are expected to exhibit some degree of
non-locality as well. From a phenomenological point of view one may
therefore be tempted to simply add to the standard form of an
amplitude equation non-local terms complying with the relevant
symmetries of the system \cite{VMMI,LBMDB}. A systematic derivation of
the amplitude equation from the basic non-local dynamics has been
accomplished in a few cases only. \cite{Elmer,LeZo,PlBu,kassner:02,NOWB}. 

In the present letter we investigate the formation of static patterns on
the free surface of a magnetic fluid subject to an external magnetic
field. In this process the non-local character of the magnetic 
interaction is of vital importance. We establish a closed, strongly non-local
equation for the free surface of the magnetic fluid and systematically
derive an amplitude equation for the surface deflection in the
vicinity of the critical field strength. Although the corresponding
linear operator is non-local the lowest order amplitude equation is
found to be {\it local} and to reproduce previous results from a
perturbative analysis of the free energy
\cite{gailitis:77,friedrichs:01}.  

Magnetic fluids are suspensions of ferromagnetic nanoparticles in 
suitable carrier liquids behaving as super-paramagnetic Newtonian
liquids~\cite{rosensweig:85}. If the strength of a magnetic field 
perpendicular to the flat free surface of a magnetic fluid exceeds a
threshold value the surface is known to become unstable due to the
normal field or Rosensweig instability \cite{cowley:67} and a
hexagonal or square pattern of fluid peaks develops
\cite{cowley:67,gailitis:77}. 

In order to keep the analysis simple it is convenient to
consider a situation with just one unstable mode analogous to the
roll solution in hydrodynamic systems. As was known experimentally
for some time \cite{barkov:77} and has been clarified theoretically
recently \cite{friedrichs:02} the corresponding ``rigde'' pattern on 
a magnetic fluid surface may be induced by an {\em oblique} magnetic
field. In this case the field component tangential to the flat surface
suppresses surface deflections in this direction.


\section{Basic equations}

\begin{figure} 
\twofigures[height=4.5cm]{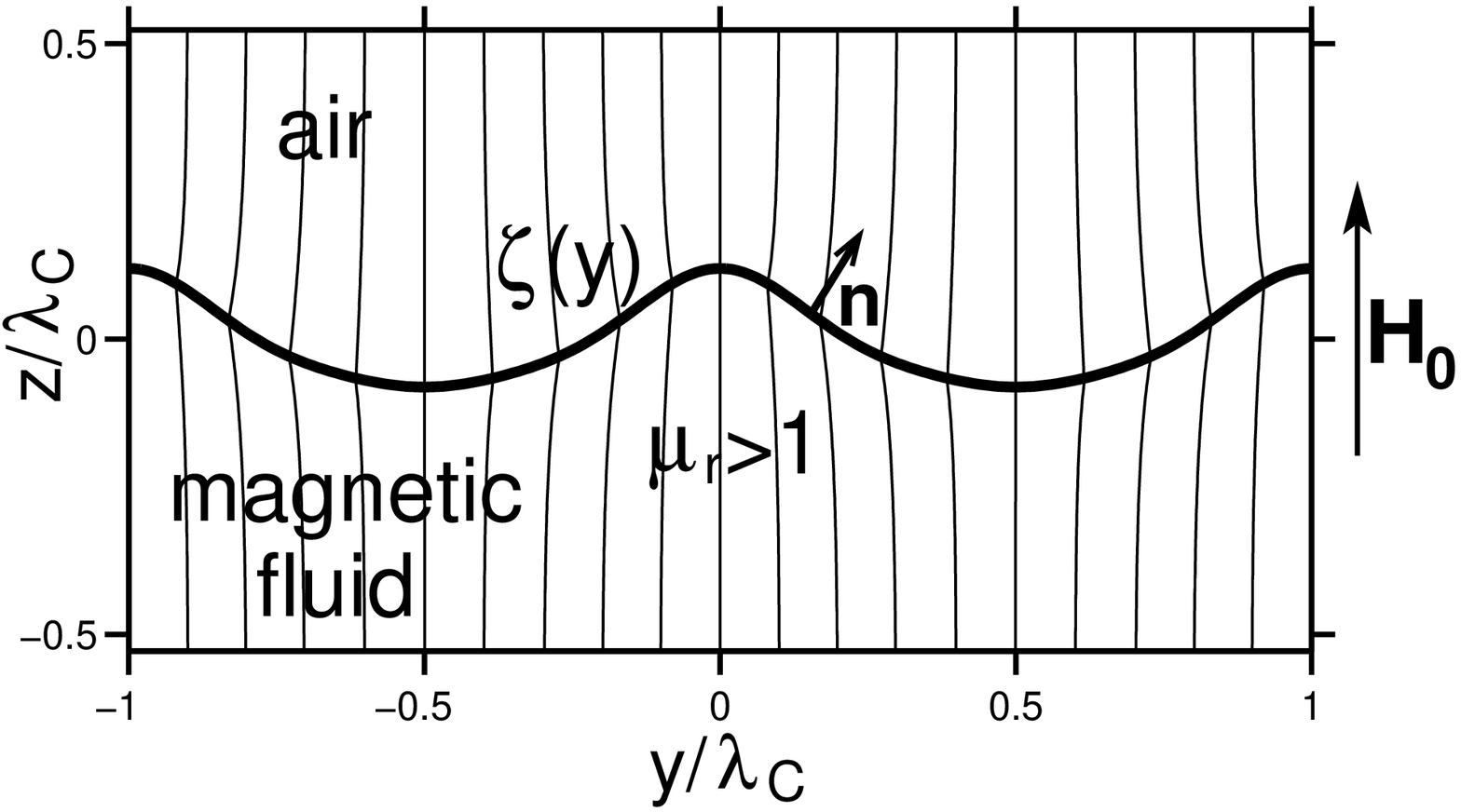}{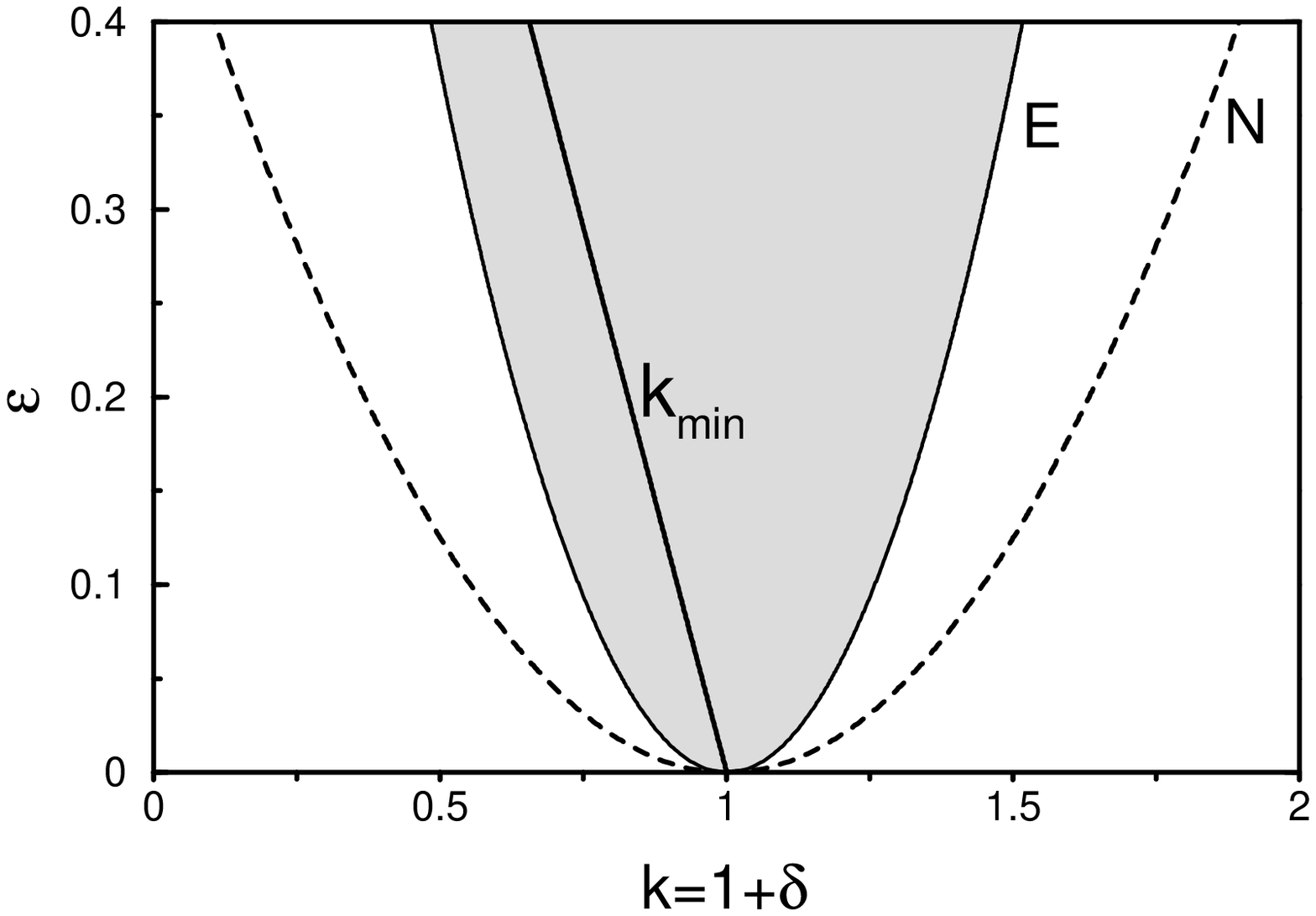}
\caption{\label{fig:1}
Static surface profile $\zeta(y)$ and magnetic field lines for a
magnetic fluid with $\mu_{\mathrm{r}}=2$ in an external magnetic field
corresponding to $\epsilon=0.0725$. The critical wavelength
$\lambda_{\mathrm{c}}=2\pi/k_{\mathrm{c}}$ is typically of the order
of 1 cm.} 
\caption{\label{fig:2}
Stability diagram for a magnetic fluid with $\mu_{\mathrm{r}}=2$ in an
overcritical magnetic field. $\eps$ is related to the magnetic field
strength by (\ref{eq:epsilon}), $\delta$ denotes the deviation from
the critical wavenumber $k_{\mathrm{c}}=1$. N is the neutral curve of
the linear instability, E denotes the Eckhaus boundary following from
the amplitude equation (\ref{eq:ampequation}). Also shown is the
wavenumber $k_{\mathrm{min}}$ minimizing the free energy as determined
in \cite{friedrichs:01}.}
\end{figure}

We consider an incompressible magnetic fluid of infinite
depth, density $\rho$ and constant permeability $\mu=\mu_0 \mu_r$ in a
magnetic field  ${\bf  H}_{0}$ which, in the absence of any 
magnetically permeable material, is homogeneous and of the form
${\bf H}_{0}=H_{\mathrm{Z}}{\bf e}_{z}+H_{\mathrm{X}}{\bf e}_{x}$.
The free surface between the fluid and the magnetically impermeable air
above is described by $z=\zeta(x,y)$ (see fig.~\ref{fig:1}). The
surface tension is denoted by $\sigma$ and gravity acts parallel to
the z-axis, ${\bf g}=-g{\bf e}_{z}$. We assume the horizontal magnetic
field component $H_{\mathrm{X}}{\bf e}_{x}$ to be strong enough to
suppress surface deflections in $x$-direction rendering the problem
effectively two dimensional. 

The static surface profile $\zeta(y)$ is then determined by the
pressure equilibrium at the surface 
\cite{landau:85,rosensweig:85} 
\begin{equation}
  \rho g \zeta - \sigma \frac{\partial_{y}^2 \zeta}
                             {[1+(\partial_{y} \zeta)^2]^{3/2}}
 -\frac{\mu_{0}(\mu_{\mathrm r}\!-\!1)}{2}
                 {\bf H}^{2} {\Big |}_{\zeta}
 -\frac{\mu_{0}(\mu_{\mathrm r}\!-\!1)^2}{2} 
                      H_{n}^{2} {\Big|}_{\zeta}= p.
\label{eq:young}
\end{equation}
Here ${\bf H}{\big |}_{\zeta}$ and $H_{n}{\big |}_{\zeta}$ denote the
magnetic field and its normal component, respectively, {\em in} the
fluid at $z=\zeta(y)$. The air pressure $p$ on the right hand-side is
assumed to be constant. The first two terms of eq.~(\ref{eq:young})
are the standard expressions describing the influence of gravity and
surface tension. The remaining terms characterize the impact of the
magnetic field on the surface profile $\zeta(y)$. 

The magnetic field ${\bf H}(y,z)$ has to fulfill the magneto-static
Maxwell equations
\begin{equation}
\nabla \cdot {\bf B} = 0 \quad \mathrm{and} \quad
\nabla \times {\bf H} = {\bf 0},
\label{eq:maxwell}
\end{equation}
where the magnetic induction 
${\bf B}(y,z) =\mu_{0}\mu_{\mathrm{r}}{\bf H}(y,z)$ is a linear 
function of the magnetic field. The field equations are supplemented
by the boundary conditions 
\begin{eqnarray}\label{eq:boundaryB}
({\bf B}^{\mathrm{(a)}}-{\bf B}^{\mathrm{(f)}}){\big |}_{\zeta} \cdot {\bf n}
&= 0 \\
({\bf H}^{\mathrm{(a)}}-{\bf H}^{\mathrm{(f)}}){\big |}_{\zeta} \times{\bf n}
&= {\bf 0},
\label{eq:boundaryH}
\end{eqnarray}
which describe the feedback of the surface profile on the magnetic
field. Here 
\begin{equation}\label{defn}
   {\bf n}=\frac{(0,-\partial_{y}\zeta,1)}
                {\sqrt{1+(\partial_{y}\zeta)^2}}
\end{equation}
is the normal vector on the free surface pointing outwards and the
upper indices $^{\mbox{(a)}}$ and $^{\mbox{(f)}}$ refer to the air
above and the fluid below the interface, respectively. The conditions 
\begin{equation}
{\bf H}(y,z) \rightarrow\left\{
  \begin{array}{rll} H_{\mathrm{Z}}\,{\bf e}_{z} &\mathrm{if} & z\to\infty\\
           H_{\mathrm{Z}}/\mu_r \,{\bf e}_{z}&\mathrm{if} & z\to -\infty
  \end{array}\right.
\label{eq:condition}
\end{equation}
ensure that the magnetic field asymptotically approaches its
externally prescribed values.

It is convenient to rescale lenght by the capillary length  
$k_{\mathrm{c}}^{-1}=\sqrt{\sigma/\rho g}$, pressure by 
$\sqrt{\rho g \sigma}$ and magnetic field by the critical
field strength of the Rosensweig instability 
\begin{equation}
H_{\mathrm{c}} = \sqrt{\frac{2\mu_{\mathrm{r}}(\mu_{\mathrm{r}}+1)
\sqrt{\rho g\sigma}}{\mu_{0}(\mu_{\mathrm{r}}-1)^{2}}}.
\label{eq:Hcinf}
\end{equation}
Moreover we introduce the supercriticality parameter
\begin{equation}
\epsilon=H_{\mathrm{Z}}^2-1
\label{eq:epsilon}
\end{equation}
and the parameter
\begin{equation}
\eta=\frac{\mu_{\mathrm r}-1}{\mu_{\mathrm r}+1}
\label{eq:eta}
\end{equation}
characterizing the magnetic susceptibility of the fluid. Denoting the
tangential component of the magnetic field by $H_t$
eq.(\ref{eq:young}) then assumes the dimensionless form   
\begin{equation}\label{eq:young2}
  \zeta - \frac{\partial_{y}^2 \zeta}
                             {[1+(\partial_{y} \zeta)^2]^{3/2}}
 -\frac{\mu_{\mathrm r}}{\eta} H_t^{2} {\Big |}_{\zeta}
 -\frac{\mu_{\mathrm r}^2}{\eta}H_n^2 {\Big|}_{\zeta}= 
           -\frac{1+\epsilon}{\eta},
\end{equation}
where the value of $p$ was determined from the reference state 
$\zeta\equiv 0$. 

The dependance of the magnetic field on the surface deflection via the
boundary conditions (\ref{eq:boundaryB}) and (\ref{eq:boundaryH}) is 
rather implicit. To make the non-locality of the problem more explicit
we take advantage to the {\em linearity} of the magnetic field
problem and employ a formal Greens function solution for the magnetic
field. To this end we write the field in the form 
\begin{equation}
  {\bf H}(y,z)= \frac{\mur-1}{2\mur} H_{\mathrm{Z}}
   \int\limits_{-\infty}^{+\infty}\! \frac{dy'}{2\pi} \,q(y')\,
      \frac{(y\!-\!y'){\bf e}_{y}+(z\!-\!\zeta(y')){\bf e}_{z}}
                       {(y\!-\!y')^2+(z\!-\!\zeta(y'))^2}\,+\,
   \frac{1+\mur}{2\mur}\,H_{\mathrm{Z}}\,{\bf e}_{z}, 
\label{eq:sources}
\end{equation}
where the so far unknown magnetic source density $q(y)$ is
nonzero at the surface $z=\zeta(y)$ only. The field (\ref{eq:sources})
already fulfills the Maxwell equations (\ref{eq:maxwell}) as well as
the boundary condition (\ref{eq:boundaryH}). The remaining boundary
condition (\ref{eq:boundaryB}) gives rise to an equation for
$q(y)$. To derive this equation we observe that the normal component
of the field (\ref{eq:sources}) is discontinuous at the interface, 
\begin{equation}
  H_n^{\mathrm{(a)}}-H_n^{\mathrm{(f)}}=
     \frac{\mur-1}{2\mur} H_{\mathrm{Z}}  
           \frac{q(y)}{\sqrt{1+(\partial_{y}\zeta)^2}}. 
\end{equation}
Moreover, interpreting the integral in (\ref{eq:sources}) for $z=\zeta(y)$
as Cauchy principal value we find for the field directly {\em at} the
surface $H_n(y,\zeta(y))=(H_n^{\mathrm{(a)}}+H_n^{\mathrm{(f)}})/2$.  
Together with (\ref{eq:boundaryB}) this gives rise to  
\begin{equation}
\frac{2\mur}{\mur+1}\;H_n(y,\zeta(y))= H_{\mathrm{Z}}
\frac{q(y)}{2\sqrt{1+(\partial_{y}\zeta(y))^2}}, 
\label{eq:relation}
\end{equation}
and using this relation in eq.~(\ref{eq:sources}) we obtain for $q(y)$ the
linear integral equation  
\begin{equation}
q=1-\eta(\partial_{y} \zeta)\mathcal{S}q+\eta\mathcal{T}q. 
\label{eq:couplep}
\end{equation}
The operators $\mathcal{S}$ and $\mathcal{T}$ act on bounded
functions $f(y)$ and are defined by 
\begin{equation}\label{defS}
(\mathcal{S}f)(y)=
\dashint\limits_{-\infty}^{+\infty} \! \frac{dy'}{\pi} \,
\frac{y-y'}{(y-y')^2+(\zeta(y)-\zeta(y'))^2} \, f(y')
\end{equation}
and
\begin{equation}\label{defT}
(\mathcal{T}f)(y)=
\dashint\limits_{-\infty}^{+\infty} \! \frac{dy'}{\pi} \,
\frac{\zeta(y)-\zeta(y')}{(y-y')^2+(\zeta(y)-\zeta(y'))^2} \, f(y').
\end{equation}
where $\int\!\!\!\!\!-$ denotes the Cauchy principal value of the
integral. Note that due to the constant term on the rhs of
(\ref{eq:couplep}) also the asymptotic conditions (\ref{eq:condition})
are fulfilled. 

With the help of (\ref{eq:sources}) we can express the magnetic field
contributions in (\ref{eq:young2}) in terms of the source density
$q(y)$ and obtain the pressure balance in the form 
\begin{equation}
  0=\zeta - \frac{\partial_{y}^2 \zeta}{[1+(\partial_{y} \zeta)^2]^{3/2}}
    +\frac{1+\epsilon}{\eta}-\frac{1+\epsilon}{1+(\partial_{y} \zeta)^2}
     \left[\frac{\left(\eta[1+(\partial_{y} \zeta)^2]\mathcal{S}q
    +(\partial_{y} \zeta)q \right)^2}{(1+\eta)\eta(1-\eta)}
    +\frac{q^2}{\eta}\right]
\label{eq:couplezeta}
\end{equation}

Eqs.(\ref{eq:couplezeta}) and (\ref{eq:couplep}) are a closed system
of integro-differential equations for the stationary surface profile
$\zeta(y)$ and the related source density $q(y)$.  

While the kernels of the operators $\mathcal{S}$ and $\mathcal{T}$ are
nonlinear functions of the surface deflection $\zeta(y)$
eq.~(\ref{eq:couplep}) is a {\em linear} equation for the source density
$q(y)$. For given $\zeta(y)$ it can hence be solved formally using the
Neumann series 
\begin{equation}
q=1+\sum_{n=1}^{\infty} \eta^n\;
      [\mathcal{T}-(\partial_{y}\zeta)\mathcal{S}]^n \, 1.
\label{eq:neumann}
\end{equation}
Replacing $q(y)$ in eq.~(\ref{eq:couplezeta}) by
this series we obtain a closed nonlinear integro-differential
equation for the free surface profile $\zeta(y)$. This equation
contains products of an infinite number of integral operators and is
per se of limited use only. Nevertheless it forms a suitable
starting point for a systematic multiple scale perturbation analysis
of the weakly non-linear regime. On the other hand this expansion can
also be performed directly on the system (\ref{eq:couplezeta})
and (\ref{eq:couplep}) which is what we do in the following. 


\section{Weakly nonlinear analysis}

We consider slightly supercritical fields $0\leq\epsilon \ll 1$ and
assume that both the surface deflection $\zeta(y)$ and the source
density $q(y)$ can be expanded in powers of $\eps^{1/2}$:
\begin{align}\label{serieszeta}
  \zeta &=\eps^{1/2}\,\zeta_1+\eps\,\zeta_2+\eps^{3/2}\,\zeta_3 +
  \dots\\ \label{seriesq}
     q &=1+\eps^{1/2}\,q_1+\eps\,q_2+\eps^{3/2}\,q_3 +  \dots .
\end{align}
For the resulting expansions of the linear integral operators
$\mathcal{S}$ and $\mathcal{T}$ it is convenient to start with the 
series representations
\begin{equation}\label{seriesS}
(\mathcal{S}f)(y)=
\sum_{n=0}^{\infty}\frac{(-1)^{n+1}}{(2n)!} \,
\dashint\limits_{-\infty}^{+\infty} \! \frac{dy'}{\pi} \,
\frac{\partial_{y'}^{2n}\left[(\zeta(y)-\zeta(y'))^{2n}f(y')\right]}
    {(y'-y)}
\end{equation}
and
\begin{equation}\label{seriesT}
(\mathcal{T}f)(y)=
\sum_{n=0}^{\infty}\frac{(-1)^{n}}{(2n+1)!} \,
\dashint\limits_{-\infty}^{+\infty} \! \frac{dy'}{\pi} \,
\frac{\partial_{y'}^{2n+1}\left[(\zeta(y)-\zeta(y'))^{2n+1}f(y')\right]}
    {(y'-y)},
\end{equation}
arising from an expansion of the denominators in (\ref{defS}) and
(\ref{defT}) and appropriate partial integration. The central building
block in these expansions is the {\em Hilbert transform} defined for
bounded functions $h(y)$ by 
\begin{equation}
(\mathcal{H}h)(y)=\dashint\limits_{-\infty}^{+\infty} \! \frac{dy'}{\pi} \,
\frac{h(y')}{(y'-y)}. 
\label{eq:hilbert}
\end{equation}
Using (\ref{serieszeta}) in eqs.(\ref{seriesS}) and (\ref{seriesT})
gives rise to expansions of $\mathcal{S}$ and $\mathcal{T}$ in powers
of $\eps^{1/2}$ which together with (\ref{serieszeta}) and
(\ref{seriesq}) are used in (\ref{eq:couplezeta}) and
(\ref{eq:couplep}). Matching powers in $\eps$ then yields a hierarchy
of linear equations for the expansion coefficients of $\zeta$ and
$q$ in the usual way. 

To order $O(\eps^{1/2})$ we find
\begin{equation}\label{eq:zetaone}
\mathcal{L}\,\zeta_1=
      \zeta_1+2\partial_y\mathcal{H}\zeta_1-\partial_y^2\,\zeta_1=0.
\end{equation}
The linear operator $\mathcal{L}$ is {\em non-local} due to the
occurence of the Hilbert transform originating from the non-local
magnetic interaction. With the help of  
$\mathcal{H}\exp(iky)=\mathrm{sgn}(k)\,i\exp(iky)$ the solution of
this equation is easily found to be 
\begin{equation}\label{eq:solutionone}
\zeta_1(y)=Ae^{iy}+A^*e^{-iy}=Ae^{iy}+ \mathrm{c.c.} \, ,
\end{equation}
with a complex valued amplitude A. As it should the first order of the
perturbation expansion hence reproduces the results of the linear
stability analysis, in particular $k_c=1$. Note also that under the
standard scalar product the operator $\mathcal{L}$ is 
{\em self-adjoint}, $\mathcal{L}^{\dag}=\mathcal{L}$. 

For $\eps>0$ not only the critical mode but a whole narrow {\em band} of
modes with $k$-values around $k_c$ are unstable. As is well known this
gives rise to long wavelength modulations of the basic
pattern \cite{cross:93}. To take account of this possibility we allow for
a slow spatial dependence of the amplitude, $A(Y)$, formalized by
the substitution 
$\partial_y \rightarrow (\partial_y + \eps^{1/2}\partial_Y)$. Using
the properties of the Fourier transform of $A(Y)$ one can also show 
$\mathcal{H}A(Y)\exp(iky)=A(Y)\mathcal{H}\exp(iky)$. 

To order $O(\eps)$ we then find 
\begin{equation}\label{eq:zetatwo}
\mathcal{L}\zeta_2=2\eta A^2 e^{2iy}+\mathrm{c.c.} \, ,
\end{equation}
with the solution
\begin{equation}
\zeta_2(y)=2\eta A^2 e^{2iy}+\mathrm{c.c.} \,.
\label{eq:solutiontwo}
\end{equation}
Here the amplitude of the solution of the homogeneous equation was set
equal to zero in accordance with the general requirement that all 
$\zeta_n$ with $n>1$ should be orthogonal to $\zeta_1$. 

Eventually, to order $O(\eps^{3/2})$ we get the inhomogeneous equation
\begin{equation}\label{eq:zetathree}
\mathcal{L}\zeta_3= 2 A\, e^{iy}+\partial_Y^2 A\, e^{iy}
           -\!\left[\frac{5}{2}-8\eta^2\right]\!\!|A|^2\! A\, e^{iy}
           +4 i \eta A\, \partial_Y \! A \, e^{2iy}
           +\!\left[\frac{1}{2}+16\eta^2\right]\!\! A^3\, e^{3iy}
           +\,\mathrm{c.c.} \, .
\end{equation}
Since $\mathcal{L}$ is singular (cf. eq.~(\ref{eq:zetaone})) the right
hand side of this equation must be orthogonal to the zero
eigenfunctions of $\mathcal{L}^{\dag}$ which are $e^{\pm iy}$. Returning
to unscaled variables we hence find the amplitude equation 
\begin{equation}\label{eq:ampequation}
0=\epsilon A+\frac{1}{2}\,\partial_y^2 A
-\!\left[\frac{5}{4}-4\eta^2\right]\!\!|A|^2\! A. 
\end{equation}
Despite the non-local character of the linear operator $\mathcal{L}$
this equation is {\em local}. The terms without derivatives are
identical with those obtained from a perturbative analysis of the free
energy of the static surface deflection
\cite{gailitis:77,friedrichs:01}. In particular we reproduce the
result that for $\eta>\sqrt{5}/4$ or equivalently
$\mu_r>3.5353$ the cubic term becomes positive and higher order
non-linear terms are needed to saturate the instability \cite{ZaSh,engel:99}. 

After imposing the solvability condition (\ref{eq:ampequation}) we can
integrate (\ref{eq:zetathree}) to get
\begin{equation}\label{eq:solutionthree}
\zeta_3(y)=4 i \eta A\, \partial_Y \! A \, e^{2iy}
+\!\left[\frac{1}{8}+4\eta^2\right]\!\! A^3\, e^{3iy}+\,\mathrm{c.c.} \, .
\end{equation}
Eqs. (\ref{eq:solutionone}), (\ref{eq:solutiontwo}), and
(\ref{eq:solutionthree}) with $\partial_Y A=0$ have been used to make 
the plot of $\zeta(y)$ shown in Fig.~\ref{fig:1}.

Building on the dispersion relation for surface waves on ferrofluids 
\cite{Abou,hwm} also a slow {\em time} dependence of the amplitude $A$
with the corresponding term on the left hand side of
(\ref{eq:ampequation}) can be included. This then allows e.g. 
to investigate the stability of solutions with slow spatial
modulations of the form 
\begin{equation}\label{eq:ansatz}
A(y)=\tilde{A}\,e^{i\delta y}.
\end{equation}
A straightforward analysis \cite{cross:93} yields as boundary for the
associated Eckhaus instability 
\begin{equation}
\delta_{\mathrm{E}}^2(\epsilon)=2\epsilon/3,
\end{equation}
which is shown in Fig.~\ref{fig:2}.


\section{Conclusion}

To determine the equilibrium shape of the free surface of a magnetic
fluid in an external magnetic field is a highly non-local
problem. Eliminating the magnetic field problem by introducing a 
suitable magnetic charge density on the free surface we have derived
a closed set of two non-local equations for the determination of the
surface profile. In a systematic multiple scale perturbation expansion
of these equations the non-locality results in a non-local linear
operator $\mathcal{L}$ involving the Hilbert transform. On the other
hand, the third order amplitude equation for the surface deflection
was shown to be local. An extension of the perturbation
expansion to other patterns as hexagons and squares or higher orders
is straightforward though tedious.

\acknowledgments
We would like to thank Harald Engel, Klaus Kassner and Chaouqi Misbah
for interesting discussions. This work was supported by the 
{\it Deutsche Forschungsgemeinschaft} under FOR 301/2-1.

\end{document}